\documentclass{icrc2009}

\usepackage{graphicx}   
\usepackage{caption}    
\usepackage[font=footnotesize]{subfig} 
\usepackage{fixltx2e}
\usepackage{url}

\newcommand{\shorttitle}[1]%
{\markboth{Proceedings of the 31\MakeLowercase{$^{st}$} ICRC, {\L}\'{o}d\'{z} 2009}{#1} }
\newcommand{\etal}{\MakeLowercase{\textit{et al. }}} 

\begin{document}

\title{Acoustic search for high-energy neutrinos in Lake Baikal:\\ status and perspectives}

\author{
\IEEEauthorblockN{
V. Aynutdinov\IEEEauthorrefmark{1},
A. Avrorin\IEEEauthorrefmark{1},
V. Balkanov\IEEEauthorrefmark{1},
I. Belolaptikov\IEEEauthorrefmark{4},
D. Bogorodsky\IEEEauthorrefmark{2},
N. Budnev\IEEEauthorrefmark{2},\\
I. Danilchenko\IEEEauthorrefmark{1},
G. Domogatsky\IEEEauthorrefmark{1},
A. Doroshenko\IEEEauthorrefmark{1},
A. Dyachok\IEEEauthorrefmark{2},
Zh.-A. Dzhilkibaev\IEEEauthorrefmark{1},\\
S. Fialkovsky\IEEEauthorrefmark{6},
O. Gaponenko\IEEEauthorrefmark{1},
K. Golubkov\IEEEauthorrefmark{4},
O. Gress\IEEEauthorrefmark{2},
T. Gress\IEEEauthorrefmark{2},
O. Grishin\IEEEauthorrefmark{2},
A. Klabukov\IEEEauthorrefmark{1},\\
A. Klimov\IEEEauthorrefmark{8},
A. Kochanov\IEEEauthorrefmark{2},
K. Konischev\IEEEauthorrefmark{4},
A. Koshechkin\IEEEauthorrefmark{1},
V. Kulepov\IEEEauthorrefmark{6},
D. Kuleshov\IEEEauthorrefmark{1},\\
L. Kuzmichev\IEEEauthorrefmark{3},
V. Lyashuk\IEEEauthorrefmark{1},
E. Middell\IEEEauthorrefmark{5},
S. Mikheyev\IEEEauthorrefmark{1},
M. Milenin\IEEEauthorrefmark{6},
R. Mirgazov\IEEEauthorrefmark{2},
E. Osipova\IEEEauthorrefmark{3},\\
G. Pan'kov\IEEEauthorrefmark{2},
L. Pan'kov\IEEEauthorrefmark{2},
A. Panfilov\IEEEauthorrefmark{1},
D. Petukhov\IEEEauthorrefmark{1},
E. Pliskovsky\IEEEauthorrefmark{4},
P. Pokhil\IEEEauthorrefmark{1},
V. Poleschuk\IEEEauthorrefmark{1},\\
E. Popova\IEEEauthorrefmark{3},
V. Prosin\IEEEauthorrefmark{3},
M. Rozanov\IEEEauthorrefmark{7},
V. Rubtzov\IEEEauthorrefmark{2},
A. Sheifler\IEEEauthorrefmark{1},
O. Suvorova\IEEEauthorrefmark{1},
A. Shirokov\IEEEauthorrefmark{3},\\
B. Shoibonov\IEEEauthorrefmark{4},
Ch. Spiering\IEEEauthorrefmark{5},
B. Tarashansky\IEEEauthorrefmark{2},
R. Wischnewski\IEEEauthorrefmark{5},
I. Yashin\IEEEauthorrefmark{3},
V. Zhukov\IEEEauthorrefmark{1}
}
\\
\IEEEauthorblockA{\IEEEauthorrefmark{1}Institute for Nuclear Research, Moscow, Russia}
\IEEEauthorblockA{\IEEEauthorrefmark{2}Applied Physics Institute of Irkutsk State University, Gagarin blvd. 20, Irkutsk 664003, Russia}
\IEEEauthorblockA{\IEEEauthorrefmark{3}Skobeltsyn Institute of Nuclear Physics  MSU, Moscow, Russia}
\IEEEauthorblockA{\IEEEauthorrefmark{4}Joint Institute for Nuclear Research, Dubna, Russia}
\IEEEauthorblockA{\IEEEauthorrefmark{5}DESY, Zeuthen, Germany}
\IEEEauthorblockA{\IEEEauthorrefmark{6}Nizhni Novgorod State Technical University, Nizhni Novgorod, 
Russia}
\IEEEauthorblockA{\IEEEauthorrefmark{7}St.Petersburg State Marine University, St.Petersburg, Russia}
\IEEEauthorblockA{\IEEEauthorrefmark{8}Kurchatov Institute, Moscow, Russia}
}

\shorttitle{V. Aynutdinov \etal Acoustic search for high-energy neutrinos}
\maketitle

\begin{abstract}
We report theoretical and experimental results of on-going feasibility studies to detect cosmic neutrinos acoustically in Lake Baikal. In order to examine ambient noise conditions and to develop respective pulse detection techniques a prototype device was created. The device is operating at a depth of 150 m at the site of the Baikal Neutrino Telescope and is capable to detect and classify acoustic signals with different shapes, as well as signals from neutrino-induced showers. 
\end{abstract}


\section{Introduction}
During the last years, neutrino astronomy is undergoing rapid development and becomes a new ``window'' to the Universe.  The large scale neutrino telescopes currently under operation (NT200+\,\cite{NT200, NT+} in Lake Baikal, AMANDA/IceCube \cite{IceCube} at the South Pole and ANTARES \cite{ANTARES} in the Mediterranean) detect the Cherenkov light emitted in water or ice by relativistic charged particles produced via neutrino interactions with matter.

Back in 1957, G.A. Askaryan has shown that a high-energy particle cascade in water, besides the Cherenkov radiation, should also produce an acoustic signal \cite{Askarian-1957}. 
The potential of the acoustic detection of particle cascades is based on the fact that the absorption length for acoustic waves with a frequency about $30$ kHz 
in sea water is at least an order of magnitude larger than that of Cherenkov radiation, in the fresh Baikal water this ratio is even close to 100 \cite{Clay}. The second fact that is favorable for the detection of acoustic signals from cascade showers at distances of hundreds of meters, or even at such long distances as several kilometers, is that the amplitude of pulses produced by showers in the near-field zone decreases only as the square root of distance, while in the far-field zone it decreases as the reciprocal of distance from the shower \cite{Askarian-1979, Learned}. Therefore, in principle, a deep-water acoustic detector of high-energy neutrinos can have a much smaller number of measuring channels compared to a Cherenkov detector with the same effective volume \cite{Askarian-1977}.

However, the technology of acoustic detection in high-energy physics is much less developed than optical methods. Since several years, however, an increasing number of feasibility studies on acoustic particle detection are performed 
[10--16].

Actually, the possibility of acoustic detection of high-energy neutrinos and the energy detection threshold are determined by the possibility of separating the signals produced by cascade showers from noise produced by other sources.\\
\begin{figure*}[ht]
\centering
\includegraphics [width=0.88\textwidth]{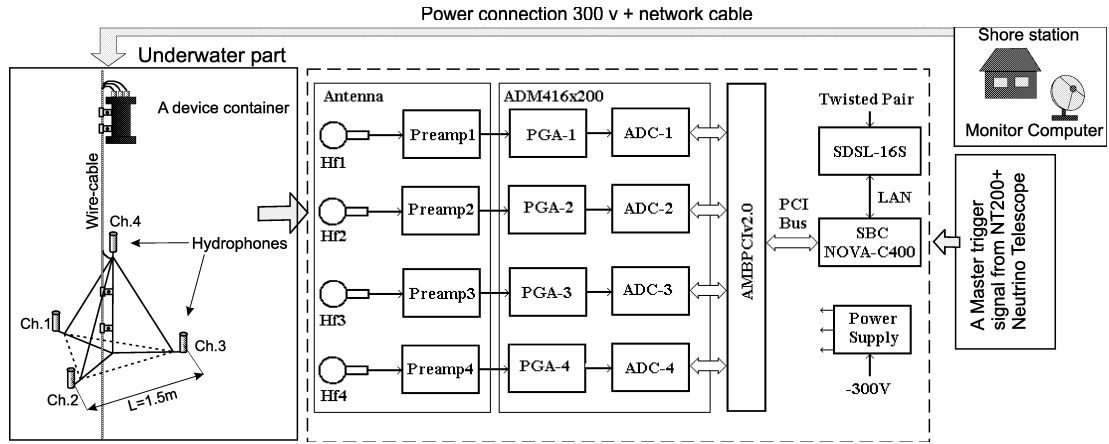}
\caption{Schematic view of the underwater $4$-channel digital device for detection of acoustic signals from high energy neutrinos.}
\label{device}
\end{figure*}

\section{Instrumentation}
\begin{figure*}[htb]
\hspace{1cm}
\includegraphics [width=0.17\textwidth, trim=0cm 0.5cm 0cm 0cm]{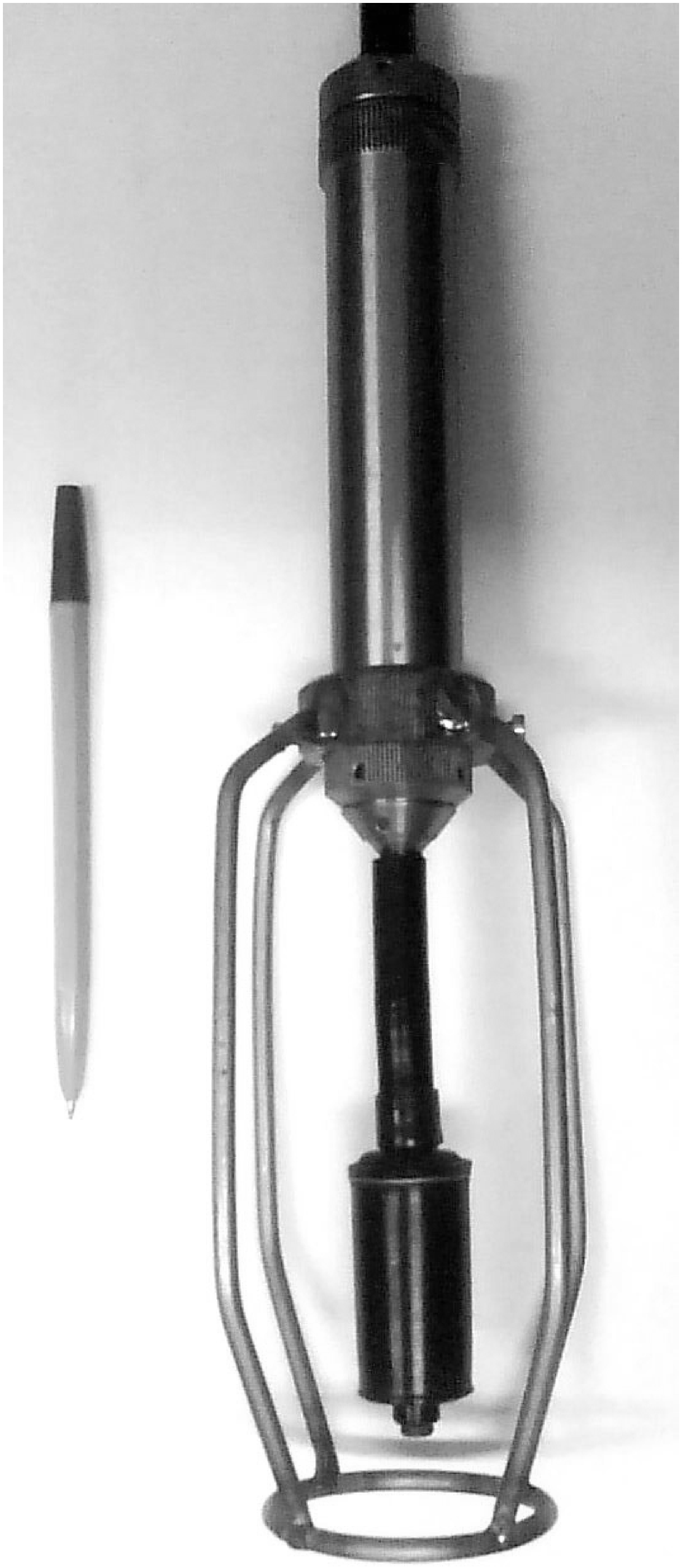}
\hfill
\includegraphics [width=0.60\textwidth, trim=0cm 0.5cm 0cm 0.3cm]{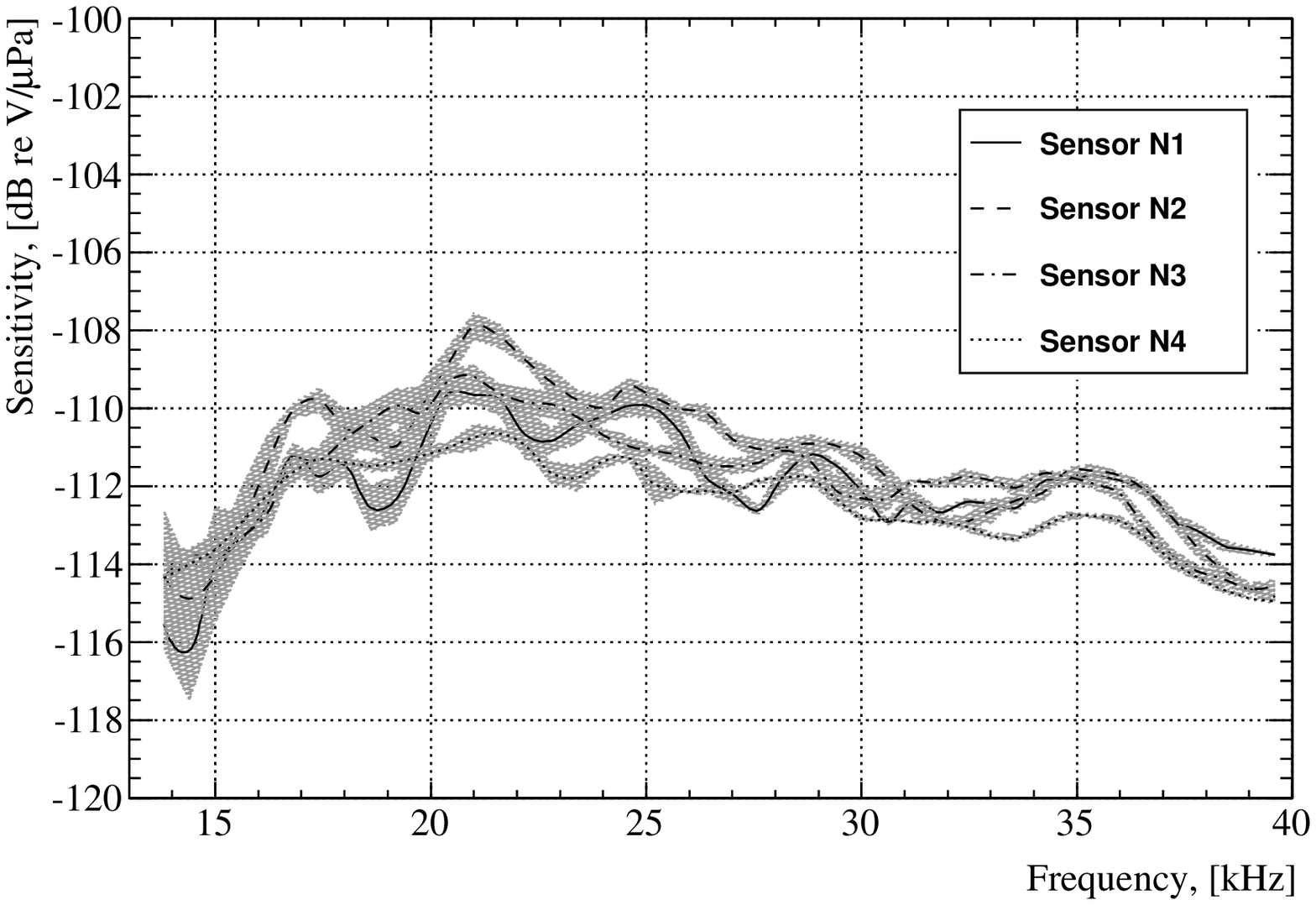}
\parbox[t]{0.35\textwidth}{\caption{Acoustic sensor with housed pre-amplifier and hydrophone H2020C.}\label{sensors1}}
\hfill
\parbox[t]{0.59\textwidth}{\caption{Sensitivity of acoustic sensors (including $70$ dB pre-amplifier gain) in units of dB V/$\mu$Pa as a function of frequency.}
\label{sensors2}}
\end{figure*}

We have constructed a digital hydro-acoustic device with four input channels, arranged on the corners of a regular tetrahedron, with edge length 1.5\,m, as shown in Fig.~\ref{device} (left). The module was designed to allow common operation with the Baikal neutrino telescope NT200+ and has been installed in April 2006 at one of the moorings of NT200+. 
Acoustic signals are recorded by four cylindrical hydrophones  H2020C (Hf-1,2,3,4 in Fig.~\ref{device} and photo in Fig.~\ref{sensors1}) 
with a sensitivity of about $1$~mV/Pa, made from a tangentially polarized piezoceramic CTS-19, 
and a maximum operating depth of about $1000$ m. All hydrophones have omni-directional pattern 
suitable for ambient noise measurements. The signals from the hydrophones are further processed by preamplifiers 
(Preamp-1,2,3,4) with $70$~dB amplification and frequency correction. 
In the range below $1$~kHz, the relative amplification is lowered by $20$ dB per octave 
in order to suppress low frequency noise. 
High frequency noise is suppressed by lowpass switched capacitor filters (LTC1569-6) following the preamplifiers. 
The further processing is performed by an IC (ADM416x200), which is mounted on the base board 
(AMBPCI v2.0) of a single board computer (NOVA-C400 Series) and includes four software programmable amplifiers (PGA-1,2,3,4) and the 16-bit over sampling {\it Analog Devices} ADC-AD7722 with a maximum conversion rate of 0.2~Msamples/sec.

The single board computer pre-processes the data and communicates with the shore computer via DSL modem and the standard neutrino telescope underwater-shore network~\cite{NANP05, RW}.
The electronics is housed in a cylindrical metallic container with $22$~cm outer diameter and $40$~cm height. 
Four hermetic connectors penetrate the upper cap of the container 
(1 - power connection 300~V; 2,3 - network twisted-pair cable, 4 - coaxial cable for trigger signal from NT200+) 
and two pairs of hydrophone connectors are arranged on opposite sides of the container. 

\section{Software and operating modes}
The module, working under Slackware Linux 11 operating system,  provides data acquisition, 
remote monitoring, and control. There are two operation modes of the instrument:
\begin{enumerate}

\item{Online search for short acoustic pulses of definite shape, 
which can be interpreted as signals from distant quasi-local sources;} 
\item{An autonomous analysis of acoustic background.}
\end{enumerate}
The device is ready for joined operation with NT200+ that could give us an opportunity to identify the properties of acoustic emission from cascades and provide energy calibration (assuming that signal strength and flux are high enough and the energy threshold low enough to collect a usable number of true coincidences). 
To reduce the amount of raw data transferred to the shore station, the pre-processing of data is done in situ using the algorithm described in \cite {ICRC29}.

\section{Calibration of acoustic sensors}
A measurement of the hydrophone sensitivity has been carried out during the winter expedition of 2009 under natural conditions (Baikal waters) by a method~\cite{Bobber} of comparing with a reference hydrophone. The hydrophone is made of the piezoceramics CTS-19 (30 mm in diameter) and calibrated in the Far Eastern Federal University (Vladivostok city). We used an impulse regime that provided measurements at shallow depths and rather small distances between emitter and receiver. The experimental setup was a virtual device, developed on the base of an industrial PC with a multifunctional measuring module (6070E) of National Instruments using a software of the same company to support the module. A measurement setup comprised two ice-holes $0.5$x$0.5$ m in size and $6.2$ m between each other. The hydrophones were mounted at a depth of $4$ m from the water surface. As a probe signal was used an impulse of varying frequency from $10$ to $40$ kHz by a step of $0.1$ kHz shaped like "square sine", with a delay between impulses of 1 second.
The frequency dependence of the emitted acoustic field was measured by the reference hydrophone, that was replaced later by non-calibrated ones. 
The obtained sensitivity curves are presented in Fig. \ref{sensors2}. The shaded bands on the curves correspond to the measurement errors.

\section{Acoustic noise characteristics at Baikal site}
Ambient noise in the ocean has been much investigated~\cite{Knudsen, Furduev}. The most intense sources of noise are waves, wind, rain, ice cracks, ship traffic, etc, which are located in the surface water layer; while biological noise and thermal noise are generated throughout the whole water column. 
The bottom of a water basin can represent a source of seismic noise and noise associated with gas seeps. 
Thermal noise represents the lower bound of underwater noise and determines the minimum 
threshold of acoustic neutrino detection~\cite{Mellen}.

We find that the qualitative behavior of noise power spectral density in Lake Baikal does follow the empirical frequency dependence typical for ocean sites~\cite{Furduev} with a slope of  spectra of about $5$--$6$ dB/octave on average (see~\cite{ACOUSTIC2} for details). In winter, the spectra are usually somewhat steeper than in summer.

\begin{figure}
\begin{center}
\includegraphics [width=0.48\textwidth, trim= 0.0cm 0cm 0.0cm 0.0cm]{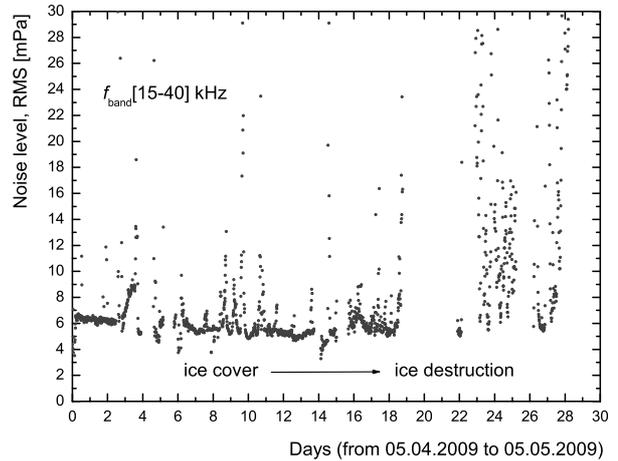}
\caption{
Acoustic noise level at the depth of $150$ m, for $1$ month of the high noise 
ice-coverage/melting period. Local time since midnight of
first measurement day.
}\label{noise_depth}
\end{center}
\vspace{-0.6cm}
\end{figure}

At any point of time the noise depends on the surface condition rather than on depth. 
Fig.~\ref{noise_depth} shows the data sample of the acoustic noise in the $15$--$40$ kHz interval,
recorded by the 4-channel device in spring 2009, i.e. during the ice-coverage and the critical ice-melting period. Periods of non-stationary conditions are visible by their high noise contribution. For open water, at stationary and homogeneous meteorological conditions the integral noise power reaches levels as low as $\sim 1$~mPa. 

\section{Pulsed noise characteristics}  
From the assumption that the main mechanism of acoustic signal generation by cascade showers is 
thermoacoustic~\cite{Askarian-1957}, it follows that the pressure pulses produced by them should 
have a bipolar shape. 
Taking into account that the high-frequency noise level in Lake Baikal can be $\sim 1$ mPa, it is possible to register showers of energy above $10^{18}$ eV 
from a distance of $100$ m, and above $10^{19}$ eV from $1000$ m~\cite{ICRC30}. At distances much larger than $1$ km, signal amplitudes of showers in Baikal water are similar or higher than in the Mediterranean due to lower sound absorption in Baikal freshwater.

\begin{figure}
\begin{center}
\includegraphics [width=0.40\textwidth, trim= 0.0cm -0.5cm 0.0cm 0cm]{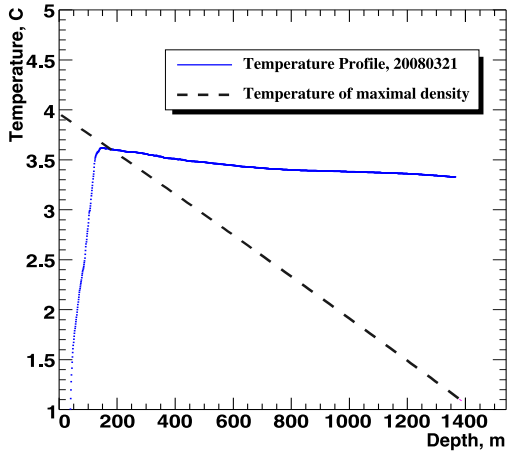}
\caption{Measured temperature as function of depth (28.3.2008) and temperature of maximal density, TMD (dashed line, calculated from \cite{Chen}).}\label{fig_temp2803}
\end{center}
\vspace{-0.5cm}
\end{figure}

The Baikal water temperature at depths below $400$~m is very stable around $\sim 3.4$-- $3.6^{\rm o}$C, see Fig.\ref{fig_temp2803} for a spring 2008 measurement. 
As also shown in the figure, the temperature is only  equal to that of maximal density
(TMD - dashed curve calculated using the formula from Ref.~\cite{Chen}) at shallow depths below $\sim$ $200$~m; at larger depths they differ significantly, since TMD falls steeply by $\sim$0.2$^{\rm o}$C per $100$~m depth.
We note, that temperature at the experimental site is permanently monitored at all depths~\cite{NT+, Budnev}.

For determining the properties of acoustic noise as the background against which neutrino induced cascades should be detected, it is necessary to measure the amplitude-time characteristics of this noise in the high-frequency spectral region.

Several years of study of acoustic noise in Lake Baikal have shown, that the background for acoustic detection 
of high-energy showers is primarily represented by pulsed noise with short pulse duration~\cite{ACOUSTIC2, ACOUSTIC1}. 

\begin{figure}
\begin{center}
\includegraphics [width=0.47\textwidth, trim= 0.0cm 0.0cm 0.0cm 0cm]{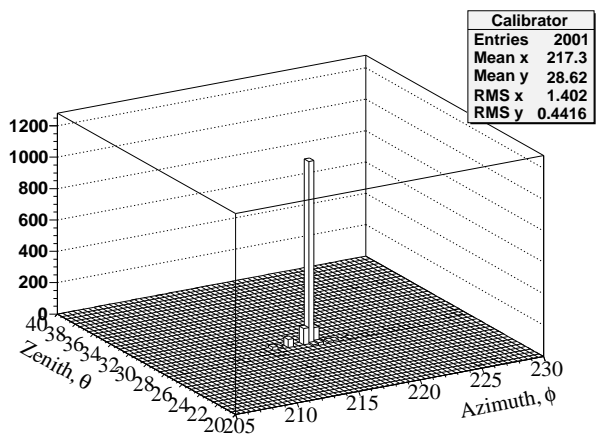}
\end{center}
\caption{Distribution of acoustic signals from calibration source versus zenith and azimuth angles.}
\label{fig4}
\end{figure}

The acoustic antenna allows us to estimate the zenith and azimuth angles of incidence of acoustic signals. In order to measure the angular resolution of the antenna, a hydrophone based calibrator of bipolar signals was developed. The calibrator was installed underwater at a distance of $50$ m from the antenna. About $5000$ acoustic signals of different shape were recorded during the measurement and processed further by a correlation analysis procedure.
Fig.~\ref{fig4} shows the distribution of reconstructed acoustic signals from calibrator versus zenith and azimuth angles. The RMS of reconstructed angles to the calibration source are $\sigma_{\phi}\approx 1.5^{\rm o}$ and $\sigma_{\theta}\approx 0.5^{\rm o}$. 

\section{Acoustic string development}
A prototype device for detection of acoustic signals from ultrahigh-energy neutrinos in water was 
constructed and is operating since April 2006 at depth of $150$ m at the site of the  Baikal Neutrino Telescope.
We suggest an acoustic detector design that is based on the deployment of a grid of such rather compact devices arranged at shallow depths ($100$--$200$ m for Lake Baikal), that mainly "watches" the water volume from the surface down to the bottom. 

The acoustic string R\&D development stage has been started this year. We plan a string which consits of three 4-channel acoustic units based on PICOe-ATOM industrial PCs, that communicate with each other via dual channel $100$/$1000$ Mbit Ethernet line. The top module on the string will have additional DSL connection to the NT200+ underwater network. Each acoustic module will be composed of directive waveguide sensors, which possess high uniform sensitivity within a cone $90$ degrees wide, offering better noise suppression to search for near-horizontal UHE neutrinos in deep water layers of the lake. For acoustic signal digitization we examine the professional Infrasonic QUARTET 4x4 channel audio cards with 24-bit/192kHz sigma delta ADCs. The test measurements with the acoustic string are scheduled for the winter expeditions of 2010/2011y. 

\section{Acknowledgments}
This work was supported by the Russian Ministry of Education and
Science, the German Ministry of Education and Research, Russian
Fund of Basic Research (grants 07-02-00791, 08-02-00198, 09-02-10001, 08-02-00432-a, 08-02-05037, 09-02-00623-a), program "Development of Scientific Potential in Higher Schools" (proj. 2.2.1.1/1483, 2.1.1/1539, 2.2.1.1/5901) and by the Grant of President of Russia NSh-4580.2006.2.


\begin{thebibliography}{00}
\bibitem{NT200}
V.~Aynutdinov et al., Astropart. Phys. 25 (2006) 140.

\bibitem{NT+}
V.~Aynutdinov et al., NIMA 602 (2009) 14.

\bibitem{IceCube}
A.~Achterberg at al., Astropart. Phys. 26 (2006) 155.

\bibitem{ANTARES}
J.~Carr [ANTARES Collaboration], NIMA 588 (2008) 80.

\bibitem{Askarian-1957} 
G.~A.~Askaryan, Atomnaya Energiya. 3 (1957) 152.

\bibitem{Clay}
C.~S.~Clay and H.~Medwin, Acoustical Oceanography (Wiley, New York), 1977.

\bibitem{Askarian-1979}
G.~Askaryan, B.~Dolgoshein, A.~Kalinovsky, NIM 164 (1979) 267.

\bibitem{Learned}
J.~G.~Learned, Phys. Rev. D 19 (1979) 3293.

\bibitem{Askarian-1977}
G.~Askaryan, B.~Dolgoshein, Pis'ma ZhETF 25 (1977) 232.

\bibitem{ARENA}
Proc. of Int. Workshop on Acoustic and Radio EeV Neutrino Dection Activities (ARENA2005), (World Sientific, Singapore), 2006, 298 p., Editors R. Nahnhauer and S. B\"oser.

\bibitem{Thompson}
Lee~F.~Thompson, NIMA 588 (2008) 155.

\bibitem{ACORNE}
S.~Danaher [ACoRNE], J. Phys. Conf. Ser. 81 (2007) 012011.

\bibitem{AMADEUS}
K.~Graf et al., J. Phys. Conf. Ser. 81 (2007) 012012.

\bibitem{NEMO}
G.~Riccobene [NEMO], J. Phys. Conf. Ser. 81 (2007) 012013.

\bibitem{SAUNDII}
N.~Kurahashi et al., Int. J. Mod. Phys.  A 21S1 (2006) 217.

\bibitem{SPATS}
S.~Boeser et al., Int. J. Mod. Phys. A 21S1 (2006) 221.

\bibitem{NANP05} V. Aynutdinov et al., 
Phys. Atom. Nuclei, V69, 11 (2006) 1914.

\bibitem{RW} V. Aynutdinov et al., NIM A567 (2006) 433.


\bibitem{ICRC29}
V.~Aynutdinov et al., in: Proc. of the 29th ICRC, Pune, India, vol. 8, 2005, p. 251.

\bibitem{Bobber}
R.~J.~Bobber, in: {\it Underwater Electroacoustic Measurements} (Washington, D.C., 1970), 360p.

\bibitem{Knudsen}
V.~O.~Knudsen, R.~S.~Alford, and J.~W.~Emling, J. Mar. Res. 7 (1948) 410.

\bibitem{Furduev}
A.~V.~Furduev, in: {\it Ocean Acoustics} (Nauka, Moscow, 1974), pp. 615-691 [in Russian]

\bibitem{Mellen}
R.~Mellen, J. Acoust. Soc. Am. 25 (1952) 478.

\bibitem{ACOUSTIC2}
V.~Aynutdinov et al., Acoust. Phys. 52 (2006) 495.

\bibitem{ICRC30}
K.~Antipin et al., in: Proc. of the 30th ICRC, Pune, India, vol. 5 (HE part 2), 2008, p. 1561.

\bibitem{Chen}
C.~T.~Chen and F.~J.~Millero,  Limnol. Oceanogr.  31 (1986) 657.

\bibitem{Budnev}
V.~Aynutdinov et al., NIMA 598 (2008) 282. 

\bibitem{ACOUSTIC1} 
V.~Aynutdinov et al., Int. J. Mod. Phys. A21S1 (2006) 117.


\end{thebibliography}
\end{document}